\documentclass[twocolumn,showpacs,preprintnumbers,amsmath,amssymb]{revtex4}

\usepackage{graphicx}
\usepackage{dcolumn}
\usepackage{bm}

\begin{document}


\title{Photo-induced spin dynamics in ferromagnetic semiconductor $p$-(Ga,Mn)As}

\author{Y. Mitsumori$^{1}$
\footnote{Present address: Communications Research Laboratory, 4-2-1 Nukui-Kitamachi, Koganei, Tokyo 184-8795, Japan.}
\footnote{Electronic address: \texttt{mitumori@crl.go.jp}}, 
A. Oiwa$^{1}$
\footnote{Present address: PRESTO, Japan Science and Technology Corporation, 4-1-8 Honcho, Kawaguchi, 332-0012, Japan.}, 
T. S\l upinski$^{2}$, H. Maruki$^{3}$, Y. Kashimura$^{1}$, F. Minami$^{3}$, and H. Munekata$^{1}$
}
\affiliation{
$^{1}$ Imaging Science and Engineering Laboratory, Tokyo Institute of Technology, 4259 Nagatsuda, Yokohama 226-8503, Japan.\\
$^{2}$ Institute of Experimental Physics, Warsaw University, Hoza 69, 00-681, Warsaw, Poland.\\
$^{3}$ Department of Physics, Tokyo Institute of Technology, 2-12-1 Oh-okayama, Meguro-ku, Tokyo 152-8551, Japan.
}%

\date{\today}

\begin{abstract}
Spin dynamics in ferromagnetic $p$-(Ga,Mn)As ($x$ = 0.011, $T_{C}$ = 30 K) has been studied by carefully comparing the decay time of the photo-induced reflectivity change with the transient behavior of polar Kerr rotation induced by photo-generated carrier spins with a femtosecond light pulse of various polarizations. As to the rising process, the rate of Kerr rotation is found comparable to the generation rate of spin-polarized carriers. For the decay process, the Kerr rotation and reflectivity signal both show the same decay rate at above the $T_{C}$, whereas, below the  $T_{C}$, the former becomes slower than the latter. The magnitude of Kerr rotation suggests that 10$^{2}$ Mn spins are revolved by injecting one hole spin. On the basis of these observations, collective rotation of ferromagnetically coupled Mn spins is discussed in terms of $p$-$d$ exchange interaction and successive transverse spin relaxation. Development of another long-lived behavior under external perpendicular magnetic fields is also disclosed.
\end{abstract}

\maketitle
\indent In III-V based ferromagnetic semiconductors \cite{Ohno99}, carrier spins, especially hole spins play an important role in that they mediate long-range ferromagnetic coupling between magnetic ions through the $p$-$d$ exchange interaction \cite{Munekata, Koshihara, Ohno00}. This mechanism opens the ways to manipulate the orientation of ferromagnetically coupled Mn spins with spin-polarized carriers generated by photons without applied magnetic fields \cite{Oiwa}. To pursue optical manipulation of magnetic properties, investigations of spin dynamics induced by optical pulses is indispensable.\\
\indent In this letter, we are concerned with spin dynamics of ferromagnetically coupled Mn spin system in ferromagnetic, $p$-type (Ga,Mn)As epitaxial layers. This has been accomplished by carefully comparing the data obtained from femtosecond time-resolved photo-induced Kerr rotation spectroscopy with decay time of photo-induced change in reflectivity. While temporal profile of photo-induced Kerr rotation coincides with that of the photo-induced reflectivity above the Curie temperature ( $T_{C}$ = 30 K), discrepancy develops between the two profiles below  $T_{C}$, as characterized by the prolonged relaxation rate in photo-induced Kerr rotation. The sign of rotation angle changes with the polarity of circularly polarized excitation light, and the rising rate of the rotation is found comparable to the generation rate of spin polarized carriers. The magnitude of the rotation indicates that 10$^{2}$ Mn spins are revolved per one hole spin. On the basis of these experimental results, ultrafast rotation of ferromagnetically coupled Mn spins is discussed. Development of a rather large, long-lived component under external perpendicular magnetic fields is also disclosed, and discussed in terms of temporal increase/decrease in hole-induced ferromagnetic coupling.\\
\indent Samples were prepared by molecular beam epitaxy (MBE) on GaAs/GaAs(100) substrates at the substrate temperature of 250 $^{\circ}$C. Thickness and Mn contents $x$ of the (Ga,Mn)As epitaxial layer are 200 nm and 0.011, respectively. Because of the lattice mismatch between the (Ga,Mn)As layer and the substrate \cite{Shen}, in-plane magnetic anisotropy is present in the (Ga,Mn)As layer. The Curie temperature, as measured by a superconducting quantum interference device (SQUID) magnetometer, is around 30 K. Hole concentration is inferred to be 10$^{20}$ cm$^{-3}$ or below, referring to the existing electrical transport data of (Ga,Mn)As with similar Mn contents \cite{Matsukura}.\\
\begin{figure}
\includegraphics[width=1\linewidth]{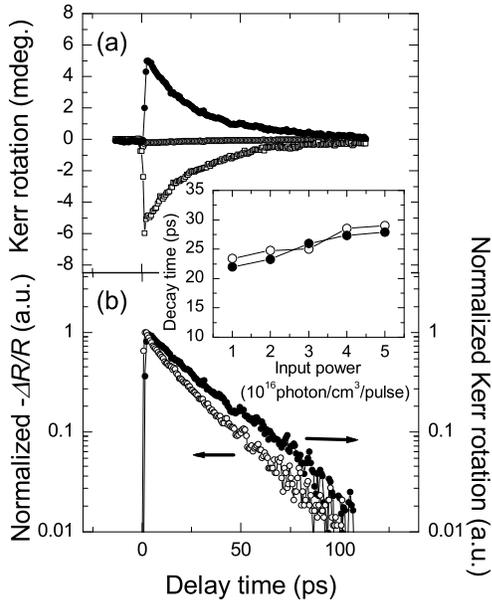}
\caption{(a) Photo-induced Kerr rotation signal at 50 K detected as a function of delay time. Excitation with three different polarization configurations, $\sigma^{+}$ (filled circles), $\sigma^{-}$  (open circles), and linear polarization (open squares) were carried out. No external magnetic field is applied. (b) Normalized photo-induced change in reflectivity signal (open circles), together with normalized photo-induced Kerr rotation signal with the $\sigma^{+}$ pump (filled circles). Excitation density of the photo-induced Kerr rotation and change in reflectivity was 3$\times$10$^{16}$ photons/pulse/cm$^{3}$. Inset shows dependence of excitation density on decay time constant of photo-induced reflectivity change (open circles) and photo-induced Kerr rotation signal (filled circles).}
\end{figure}
\begin{figure}
\includegraphics[width=1\linewidth]{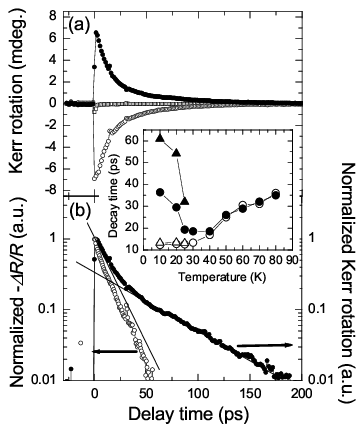}
\caption{(a) Photo-induced Kerr rotation signal at 20 K detected as a function of delay time. Excitation with three different polarization configurations, $\sigma^{+}$ (filled circles), $\sigma^{-}$ (open circles), and linear polarization (open squares) were carried out. No external magnetic field is applied. (b) Normalized photo-induced reflectivity change (open circles), together with normalized photo-induced Kerr rotation signal for $\sigma^{+}$ pump (filled circles). Inset: Temperature dependence of decay time constants extracted from temporal profiles of photo-induced Kerr rotation (filled circles) and photo-induced reflectivity change (open circles), assuming a single exponential decay profile. Open and filled triangles represent decay time constants extracted by bi-exponential curve fitting (solid thin lines). Excitation density of all measurements was 3$\times$10$^{16}$ photons/pulse/cm$^{3}$.}
\end{figure}
\begin{figure}
\includegraphics[width=1\linewidth]{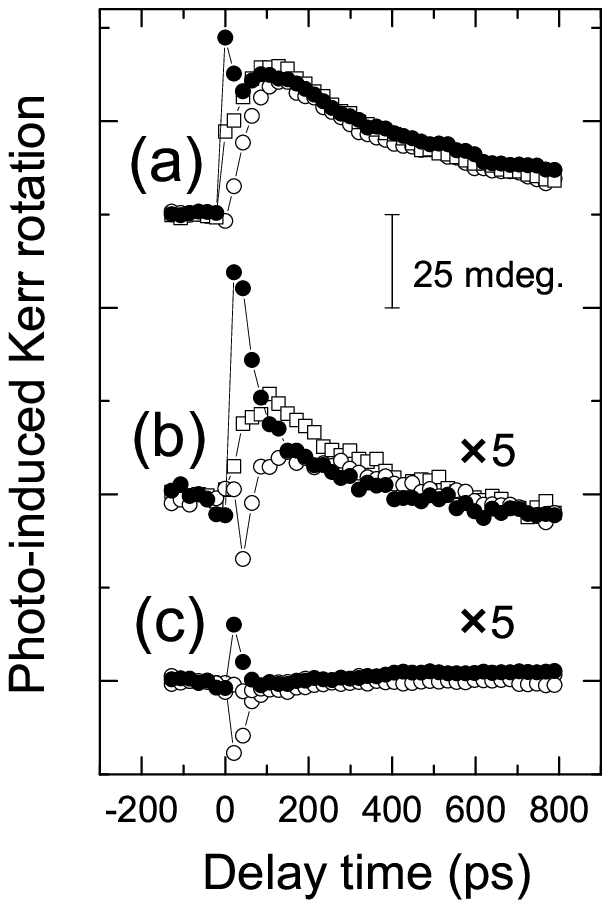}
\caption{Temporal profiles of photo-induced Kerr rotation signal measured with the magnetic field of 1T applied normal to the sample plane for (a) 20 K, (b) 50 K and (c) 70 K. Data represented by filled circles, open circles, and open squares are those obtained by excitations with $\sigma^{+}$, $\sigma^{-}$, and linear polarization, respectively. Excitation density of all measurements was 3$\times$10$^{16}$ photons/pulse/cm$^{3}$.}
\end{figure}
\indent Time-resolved polar Kerr rotation measurements, by means of pump-and-probe technique, were carried out to study spin dynamics. Light source was a cw mode-locked Ti:Sapphire laser pumped by a cw frequency-doubled Nd:YAG laser. Duration and repetition rate of laser pulses were $\sim$120 fs and 78 MHz, respectively. The laser pulse, whose central photon energy was tuned at 1.579 eV, was split in pump and probe pulses by a beam-splitter. The value of photon energy was slightly higher than the band gap of (Ga,Mn)As, and was appropriate to generate low-energy electrons and holes in the samples. Intensity of the pump pulse impinging on samples was varied between 1 and 5$\times$10$^{16}$ photons/pulse/cm$^{3}$. Polarization of pump pulses was controlled by a quarter-wave plate to be either in a linear or right- ($\sigma^{+}$) / left- ($\sigma^{-}$) handed circular polarization so that spin de-polarized or polarized carriers was respectively generated in the samples. The probe pulse of 1$\times$10$^{15}$ photons/pulse/cm$^{3}$ was fixed in a linear polarization so that the information as to spin dynamics was represented as a pump-induced change in Kerr rotation angle of the probe pulse. The change in the rotation angle was measured by means of the balancing detection technique with an accuracy of 10$^{-4}$ deg.  Dynamics of photo-generated carrier was studied by the pump-induced change in reflectivity. In this experiment, polarization of pump and probe pulses were both in linear polarization. Both time-resolved Kerr rotation and reflectivity experiments were carried out at temperatures ranging from 10 to 80 K, either under the applied magnetic field of 0 or 1 Tesla along the direction normal to the sample plane (Faraday configuration).\\
\indent We first start with time-resolved photo-induced change in reflectivity and Kerr rotation above  $T_{C}$. As shown in Fig.1(b), the reflectivity signal rises steeply at zero picosecond (ps) and decays rapidly. From the single exponential decay curve, the decay time of the reflectivity signal is estimated to be 25$\pm$1 ps. This fast decay time is most likely attributed to the crystal defects formed during the low-temperature (LT) epitaxial growth. It is known that LT-GaAs and Be doped LT-GaAs ($p$-LT-GaAs:Be) epitaxial layers contain excess ionized As and associated defects that work as electron and hole traps, which give rise to the reflectivity and differential transmission signal of sub-picosecond \cite{Gupta, Loch, Hamil}. The present (Ga,Mn)As is heavily $p$-type, so that hole traps are most likely saturated. On the other hand, the electron traps are not saturated, especially under the present low excitation density. In fact, the decay time of the reflectivity signal is independent on the excitation density, as shown in the inset. Consequently, the observed reflectivity signal mainly reflects the initial trapping time of the photo-generated electron \cite{Loch, Hamil}. This suggests that photo-generated holes stay longer in the valence band than the decay time of the reflectivity.\\
\indent Turning eyes on the behavior of time-resolved Kerr rotation signals (Fig.1(a)), the temporal profile is very similar to that observed for the reflectivity change. The rotation angle is symmetric with respective to the circular polarization, being positive and negative for the excitation of $\sigma^{+}$ and $\sigma^{-}$, respectively. For the linearly polarized excitation, the rotation signal vanishes. These results indicate that the Kerr rotation profile at 50 K is governed by spin relaxation of carriers generated by the circularly polarized pump pulse. Most likely, it is free electrons, but not holes, that are primarily responsible for the Kerr signal, since spin relaxation time of holes is known to be very fast, being less than a few ps in GaAs \cite{Damen, Hilton}. Our inference is based on the assumption that band structure of (Ga,Mn)As at above the Curie temperature can be approximated with that of a non-magnetic GaAs, at least in the vicinity of $\Gamma$ point.\\
\indent Below  $T_{C}$, the decay of Kerr rotation signal becomes clearly slower than that of reflectivity signal, as shown in the data taken at $T$ = 20 K (Figs.2(a) and (b)). As to the dependence of polarization of pump pulse on the sign of Kerr rotation angle, the relationship observed at temperatures above  $T_{C}$ holds also below  $T_{C}$ for both fast and slow components. Looking the data carefully, we notice that a slower component is superimposed on a fast decay component. A single exponential fit of the Kerr rotation signal results in the overall decay time constant of 30 ps. More precise analysis with bi-exponential curve fitting $[A_{1}\exp(-t/\tau_{1})+A_{2}\exp(-t/\tau_{2})]$ yields $\tau_{1}$ = 13 ps for the fast component and $\tau_{2}$ = 54 ps for the relatively slow component. The fast component coincides with the photo-generated electron trapping time obtained from the reflectivity data. The contribution of the relatively slow component becomes smaller with increasing temperature. The application of the bi-exponential fit became rather difficult for data taken at 30 K and higher. With single exponential fit, however, we were able to notice the presence of the relatively slow component up to around the Curie temperature, as shown in the inset in Fig. 2. As a whole, above  $T_{C}$, spin relaxation occurs with the same time constant as that of the electron trapping, whereas, below  $T_{C}$, spin relaxation slows down. This clearly indicates that the development of long-range ferromagnetic alignment of Mn spins, accompanied by the spin-polarized-hole bands, seems to be responsible for the occurrence of relatively slow decay.\\ 
\indent Within hole spin coherence time, photo-generated hole spins are aligned perpendicular to the sample plane and interact with ferromagnetically coupled Mn ions through the $p$-$d$ exchange interaction. This gives rise to the generation of a perpendicular effective magnetic field, and results in immediate rotation of Mn spins from lateral to perpendicular direction. Comparing the magnitude of Kerr rotation at the peak of photo-induced rise with the value of Kerr rotation angle at saturation magnetization, the numbers of rotated Mn is estimated to be about 10$^{18}$ cm$^{-3}$ per excitation pulse; namely, one hole spin rotates about 10$^{2}$ of Mn spins, assuming the quantum efficiency of photo-carriers to be unity. This fact strongly suggests the occurrence of collective rotation of ferromagnetically-coupled Mn spins, which could be a new type of excitation. In the previously reported cw-excitation experiments \cite{Oiwa}, the number of rotated Mn spins was deduced to be 10$^{6-7}$ cm$^{-3}$ per one hole spin. In the cw-experiments, continuous generation of spin-polarized hole makes it possible to accumulate hole spins. This results in the enhanced numbers of rotated Mn spins.\\
\indent The relaxation process of the rotated Mn ion spins can be viewed as transverse spin relaxation through the Larmor precession. The magnetic field that drives the precession comes from in-plane magnetic anisotropy of 0.2 Tesla, as estimated from the difference in $MH$ product between in-plane and out-of-plane magnetization data \cite{Ani}. This field, with a gyromagnetic constant of electron $\gamma$ = 2.2$\times$10$^{5}$ m/As, gives rise to the oscillation time of about 200 ps, being not so far from the decay time constant obtained experimentally. The absence of oscillatory behavior in Kerr rotation signal may either be due to free induction decay or homogeneous relaxation. Knowing that the decay profile is not Gaussian but exponentional, we are able to conclude that the homogeneous relaxation is the predominant process. In other words, there is a strong damping factor being present in the Larmor precession in the present case. The origin of damping may be phonon scatterings.\\
\indent Another interesting behavior has been found in the situation where initial spin states are aligned perpendicular to the sample plane. Examples are shown in Fig. 3 by the data taken with $H_{\perp}$ = 1T at which magnetization is saturated along the direction normal to the sample plane. A new component, being characterized by slowly-increasing ($\tau$ = 150 ps) and decreasing ($\tau$ = 500 ps) behaviors, appears below  $T_{C}$ ($T$ = 20 K, curve \lq a') in addition with the fast-decaying component. The magnitude of induced Kerr rotation at the maximum (40 mdeg) is fairly large, being comparable to 10\% of the saturation at 20 K. Furthermore, the long-lived Kerr rotation signal is independent on the polarization of pump pulse, suggesting that it is not a spin relaxation process but an energy relaxation process that is taking place. Here, we should recall that the lifetime of holes is long because of the predominance of electron trapping over hole trapping in the $p$-type sample. Mn states, if they can weakly bind holes \cite{Hirakawa, Takahashi}, would also contribute to increasing the hole lifetime. Within the lifetime of photo-generated holes, ferromagnetic coupling is enhanced, making Mn-spin subsystem possible to further align parallel at the given finite temperature. This is a relatively slow process, because the change in relative alignment between Mn spins is always accompanied by the changes in energy and momentum of hole-spin subsystem through emission and absorption of phonons. Compared with the complicated rising process, the decaying process is more straightforward; we believe that this process is dominated by the recombination between holes and trapped electrons.\\
\indent The long-lived component is still observable above the Curie temperature, as seen in the data taken at 50 K (curve \lq b'). Hole injection causes the variation of the Curie temperature to a slightly higher temperature, which results in enhancement in the paramagnetic susceptibility at higher temperature side near  $T_{C}$. The effect vanishes completely at 70 K (curve \lq c') which is far from  $T_{C}$. The long-lived component is also observable in ferromagnetic samples with different Mn contents of 0.068 (data not shown). In this case, the rise and decay time constants are much longer than the $x$ = 0.011 sample. In summary, our experimental results show that optical manipulation of ferromagnetically coupled Mn spin system is possible in various time scales with III-V-based magnetic alloy semiconductors.\\
\indent This work was started at \lq\lq Photomagnetic semiconductor" project (1999-2002) of \textit{Kanagawa Academy of Science and Technology}, Kanagawa, Japan, and has been being continued with partial support from Scientific Research in Priority Areas \lq\lq Semiconductor Nanospintronics" (2002 - ) of The Ministry of Education, Culture, Sports, Science and Technology, Japan.\\


\newpage


\begin{thebibliography}{17}
\bibitem{Ani} Anisotropy field $H_{a}$ is estimated by $H_{a}$ = $2K/M_{s}$, where $K$ and $M_{s}$ are anisotropy constant and saturation magnetization, respectively. $K$ was estimated from the difference in $MH$ products between in-plane and out-of plane magnetization curves of the $x$ =0.011 sample.
\end{thebibliography}
\end{document}